\documentclass[%
 aip,
 amsmath,amssymb,
 reprint,%
]{revtex4-1}

\usepackage{bm}
\usepackage{natbib}
\usepackage{graphicx}

\begin{document}
\textfloatsep 10pt

% Author Orchid ID: enter ID or remove command
\newcommand{\orcidauthorA}{0000-0002-3912-4727} % Add \orcidA{} behind the author's name

\title{Wave-controlled bacterial attachment and formation of biofilms }

% Authors, for the paper (add full first names)
\author{Sung-Ha Hong, Jean-Baptiste Gorce, Horst Punzmann, Nicolas Francois, Michael Shats and Hua Xia}
%\author{ }%\orcidA{}}
\email{hua.xia@anu.edu.au}

\affiliation{Research School of Physics, The Australian National University, Canberra ACT 2601, Australia}

\date{\today}

\begin{abstract}
The formation of bacterial biofilms on solid surfaces within a fluid starts when bacteria attach to the substrate. Understanding environmental factors affecting the attachment and the early stages of the biofilm development as well as the development of active methods of biofilm control are crucial for many applications. Here we show that biofilm formation is strongly affected by the hydrodynamics of flows generated by surface waves in layers of bacterial suspensions. Deterministic wave patterns promote the growth of patterned biofilms while wave-driven turbulent motion destroys the patterns. The location of the attached bacteria on a solid substrate differs from the settlement location of inactive bacteria and of the passive micro-particles: strong biofilms form under the wave antinodes while passive particles and inactive bacteria settle under nodal points.  By controlling the wave lengths and horizontal mobility of the wave patterns, one can either shape the biofilm formation and enhance the biofilm growth, or discourage the formation of biofilm patterns. The results suggest that the deterministic wave-driven transport, rather than hydrodynamic forces, determine the preferred location for the bacterial attachment. 
\end{abstract}

\maketitle

\section{Introduction}

The motion of particles on the fluid surface perturbed by waves is a challenging theoretical and mathematical problem with only a handful of exact solutions. The prediction of transport of particles on and under such surfaces has numerous applications, from spreading of the pollutants to clustering and settlement of living microorganisms. In recent years, experimental progress on understanding wave-driven particle motion has been impressive. Faraday waves, or parametrically excited surface waves \cite{Faraday} have been well characterised and understood due to a large body of laboratory experiments and numerical simulations. At low excitation amplitude, these nonlinearly generated waves form stable patterns \cite{Douady1990, Binks1997, Kudrolli1996, Arbel1998, Zhang} which, as the forcing is increased, become unstable and eventually create turbulent motion of fluid in the horizontal direction \cite{Goldman,Fineberg}. Such turbulence reproduces in detail the statistics expected in ideal two-dimensional turbulence \cite{von Kameke, Francois2013,Francois2014, Xia2013, Xia2014}. The description of disordered Faraday waves as ensembles of quasi-particles, or oscillons \cite{Lioubashevski1999, ShatsPRL2012,Xia2012} allows to characterise transitions from linear regime to turbulence, via the introduction of horizontal mobility of oscillons. The oscillon motion turns into a random walk which determines the diffusive transport of fluid particles on the surface \cite{Francois2015}. Since waves also generate fluid motion below free surfaces, such motion also affects transport of particles and sedimentation near the solid bottom in thin layers. One would expect that the wave-driven transport should be essential for the process of settlement and attachment of microorganisms, such as bacteria, during early stages of the formation of biofilms.

Though many bacterial species are very efficient swimmers, they prefer to colonise submerged surfaces by building biofilms which are self-produced aggregates of microorganisms in which bacteria are embedded in a complex three-dimensional matrix of extracellular polymeric substances. While bacterial biofilms are often associated with their adverse effects on surrounding organisms or hosts (medical implant surfaces, water pipes, etc.), many bacteria are beneficial for a variety of environmental, engineering and medical applications such as water treatment and biotechnologies aiming at creation of new materials \cite{Costerton1995, Costerton1999, Flemming2002,Singh2006, Wuertz2003,Grady2011, Nguyen2014, WoodPNAS2016}. Some applications will benefit from the ability to control and shape the growth of biofilms, for example for the development of bacterial scaffolds for tissue engineering \cite{Bottan2015,Serpooshan2017} or for growing bacterial cellulose which is a structural component of some biofilms \cite{Svensson2005}. 

The microbial consortia can be shaped using different patterning techniques. Biofilm lithography has been used to pattern biofilms using optically controlled microbial gene expression. Patterned substrate modification together with specific antibodies were used to immobilise and pattern live bacterial cells \cite{Suo2008}. Communities of different bacterial species have been constructed using microfluidic devices to control spatial structure and chemical communication \cite{KimPNAS2008}. Bacteria are also sensitive to physical stimuli and mechanical cues such as hydrodynamic forces, adhesive forces, and the rheology of their surroundings \cite{Stone2015, Raspaud2017, Stone2018,Stone2018_2}. For example, it was shown recently that fluid flows control the microscopic structure and three-dimensional morphology of biofilms \cite{DrescherNP2019}. Other mechanical factors, such as mechanical vibration of the substrate, also affect bacterial attachment leaving a footprint on the biofilms \cite{Murphy2016}. 

\begin{figure*}
\includegraphics[width=16cm]{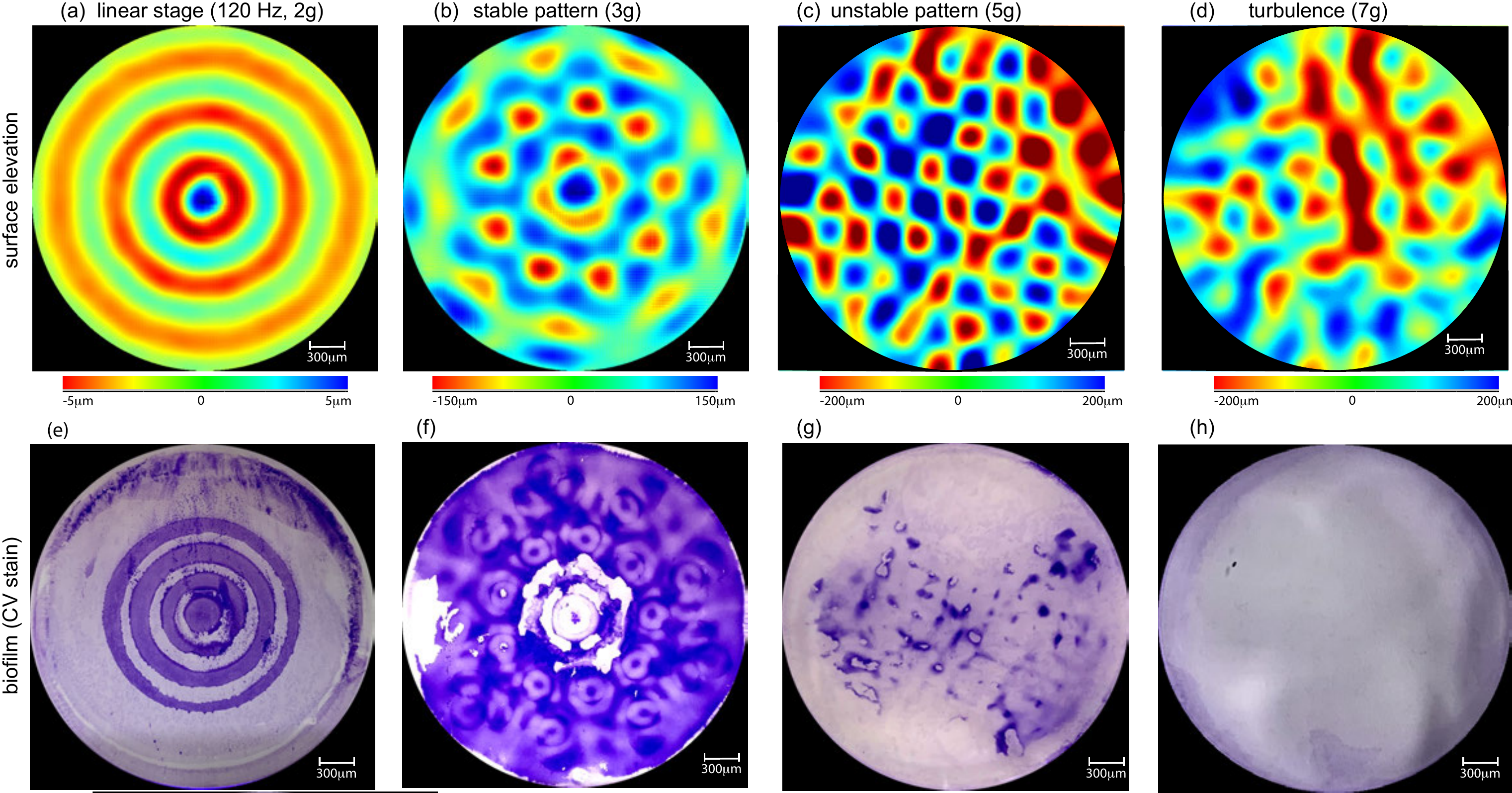}
\caption{\label{Fig_wave_biofilm}  (a-d): Measured instantaneous contour plots of the surface elevation. Red/blue colors correspond to peaks/troughs of the waves. (e-h): corresponding images of the crystal violet stain of the biofilm attached to the bottom of the fluid cell after exposure of the bacterial suspension to the waves for 24 hours. Darker colours correspond to thicker biofilms. Fluid cells are vibrated at $f_s=120$ Hz at the vertical accelerations of (left to right) $a= 2g, 3g, 5g$ and $7g$.}
\end{figure*}

The motion of a liquid must play important roles in the process of bacterial attachment to the solid substrate and on the biofilm development. On one hand, hydrodynamic forces may encourage the settlement of bacteria at particular spots. On the other hand, the flows may create favourable conditions for the biofilm development by establishing the delivery of nutrients, oxygen or other essential components to the biofilm location. Recent progress in understanding the motion of particles in fluids whose surface is perturbed by hydrodynamic waves offers new ideas on how such flows can be used to control and shape the formation of biofilms.

Here we report the development of bacterial biofilms at the bottom of vertically vibrated containers. To uncover factors affecting bacterial attachment, we image surface waves, visualise sedimentation of passive microparticles and inactive bacteria, vertical mixing by the surface waves, and study biofilms developing at the bottom. 

We find that structured, deterministic wave-driven flows encourage the development of strong biofilms attached to the bottom of the microplates. Such biofilms are characterised by regular periodic patterns which are correlated with the patterns of the surface waves. Biofilms are formed under the wave antinodes (the peak-trough locations) while inactive bacteria and passive particles accumulate under nodal points where the surface elevation is constant in time. The biofilm patterns are scalable: the characteristic spatial period of a pattern can be adjusted by changing the wave frequency and its amplitude. At higher wave amplitudes, the wave field becomes disordered, leading to turbulent transport and intense mixing in the fluid. In this regime, the biofilm mass is reduced and a pattern does not develop.

\section{Results}

\subsection{Biofilm formation under surface waves}
Surface waves induce motion of fluid particles on the surface of a liquid and in the bulk. For linear small amplitude waves, the velocities of fluid particles exponentially decrease as a function of the distance from the surface. The depth of the liquid is an important factor affecting both the wave dispersion relation \cite{FluidMechanics} and trajectories of particles settling at the bottom, and thus, the sedimentation efficiency \cite{Douady1990, Saylor2005, Perinet2017}. The dispersion relation of waves in the finite-depth fluid layer is given by $\omega^2=(gk+\frac{\sigma}{\rho}k^3)\tanh (kh)$, where $\omega$ is the wave frequency, $h$ is the layer depth, $k=2\pi/\lambda$ is the wavenumber, $\sigma$ is the surface tension coefficient, and $g$ is the acceleration of gravity. If a  fluid container is vertically vibrated at the frequency $f_s$ with sinusoidal acceleration above a certain threshold, Faraday waves at the frequency of $f_F=f_s/2$ are excited at the liquid surface. Here we investigate the biofilm formation in the presence of Faraday waves excited at $f_s$ from 45 to 120 Hz in a broad range of vertical accelerations. Detailed studies are performed at $f_s=120$ Hz ($f_F=60$ Hz) which corresponds to the Faraday wave length of $\lambda_F \approx 5$ mm. The layer thickness of the solution is kept at $h=2$ mm for all the experiments to allow waves to affect the motion of the fluid at the bottom. Thicker layers of fluids do not show strong patterning of biofilms. 

Figure~\ref{Fig_wave_biofilm} shows the results from four experiments in which a fluid cell filled with the bacterial suspension of \textit{Escherichia coli}  is vertically vibrated at the frequency of 120 Hz at different vertical accelerations $a$. The wave patterns, measured using synthetic Schlieren technique \cite{Moisy2009} (see Methods), show circular wave fronts for standing waves at $a=2g$, stable flower-shaped pattern at $a=3g$, unstable but ordered square pattern at $a=5g$, and a constantly evolving in time turbulent wave field at $a=7g$. Biofilms that develop during 24 hours exposure to such waves are illustrated in the bottom row. The biofilms (visualised using crystal violet stain) have the spatial structure similar to that of the Faraday waves at lower accelerations of $a=(2-5)g$, while in the turbulent wave field at $a=7g$ no clear biofilm pattern is observed.

\begin{figure}
\includegraphics[width=8cm]{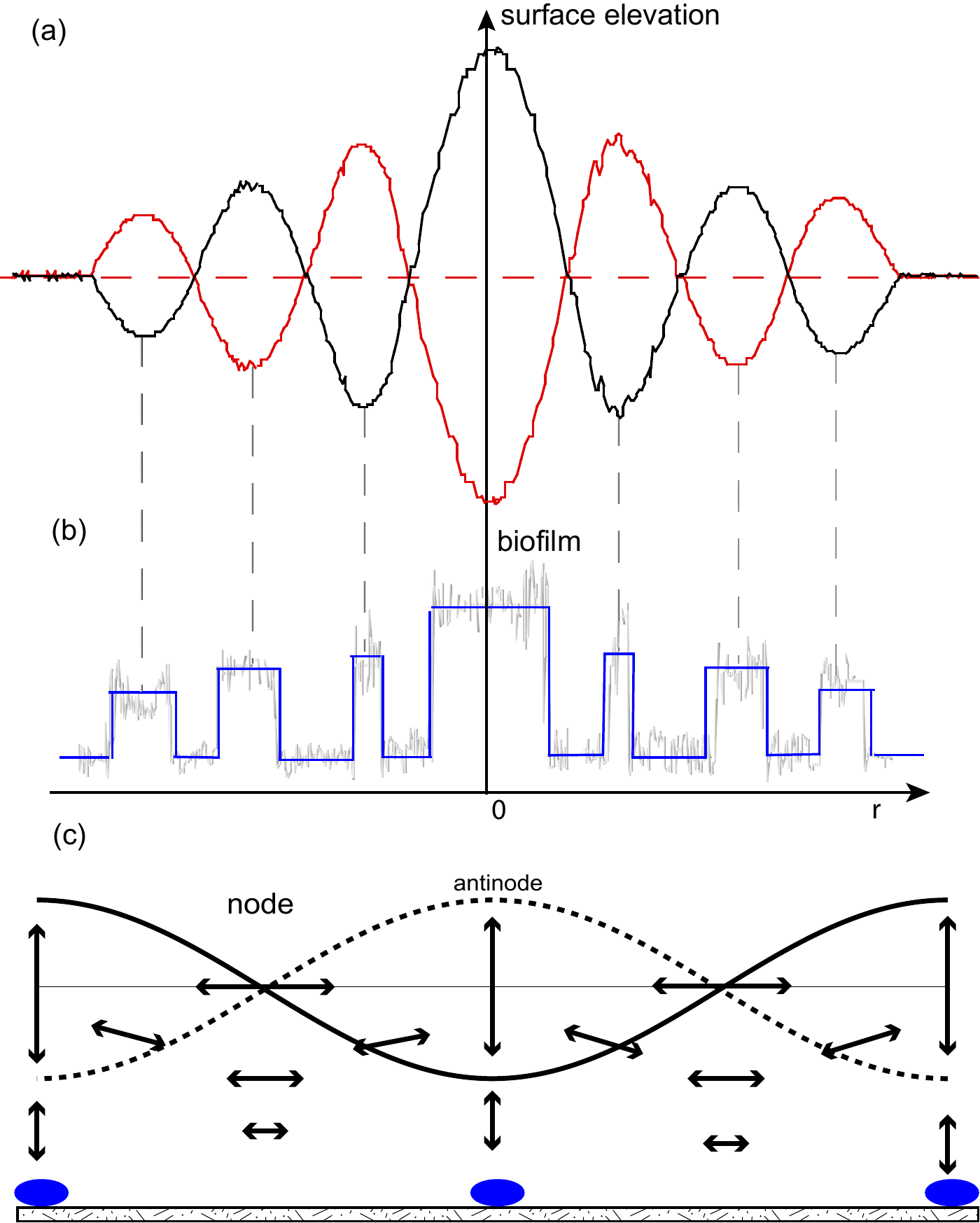}
\caption{\label{Fig_2g} (a) A profile of the surface elevation produced by the Faraday wave at the vertical acceleration of $a=2g$, and (b) corresponding profile of a crystal violet density (approximately proportional to the biofilm thickness). (c) A schematic of the fluid motion under standing surface waves \cite{Wallet1950}. Blue areas at the bottom indicate the locations of the biofilm growth.}
\end{figure}

The comparison between the biofilm strength (derived from the crystal violet intensity) and the amplitude of the surface elevation, Fig.~\ref{Fig_2g}, shows that thicker biofilms form under the locations of the wave antinodes. Note that in Fig.~\ref{Fig_wave_biofilm} the number of rings in the biofilm at $a=2g$ is twice the number of the wave periods of the wave. The biofilm pattern corresponds to two antinodes per wave period. The fluid motion at the antinodes in a standing wave is vertical, while at the positions of the wave nodes, the fluid particles move almost horizontally, as illustrated in the schematic of Fig.~\ref{Fig_2g}(c). As discussed below, the position of the biofilm under standing waves does not coincide with the expected locations of the sediment of the passive microparticles. Generally, a higher wave amplitude at a particular location leads to a thicker biofilm underneath it, as seen in Fig.~\ref{Fig_2g}(a,b).

\begin{figure}
\includegraphics[width=8cm]{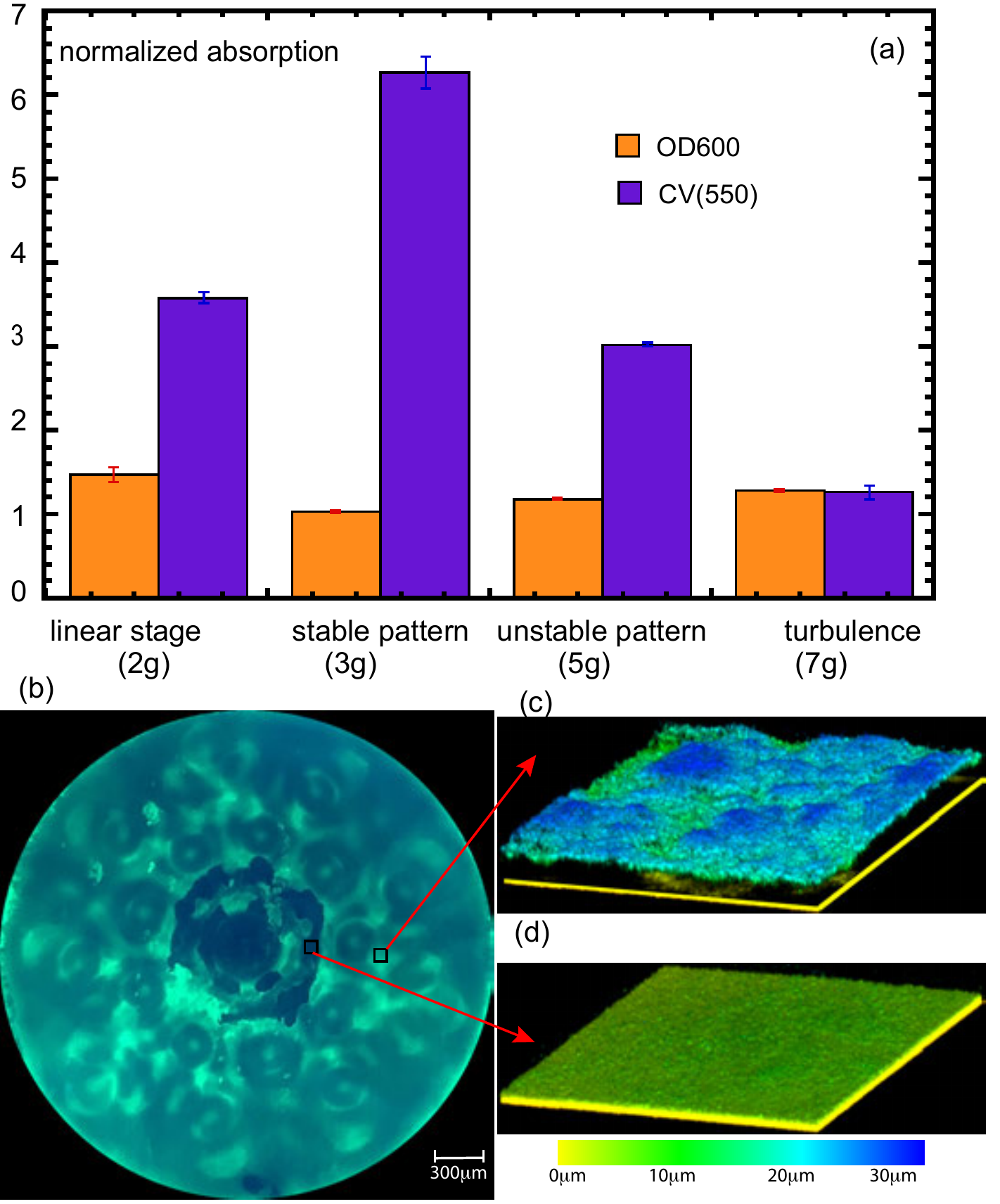}
\caption{\label{Fig_confocal} (a): Total mass of the biofilms (measured as the normalised absorption, or the optical density of the crystal violet stain at 550 nm) generated in a microplate well (35 mm diameter) at four different vertical accelerations (the vibration frequency is 120 Hz) - violet bars. The optical density of the bacterial suspension was approximately the same (measured at 600 nm) as shown by the orange bars.
(b): A fluorescent image of the biofilm formed in a well exposed to vibration at the vertical acceleration of 3$g$. (c, d): Reconstructed 3D surface of the biofilm using the Confocal Laser Scanning Microscopy at two regions of interest. }
\end{figure}

\begin{figure*}
\includegraphics[width=16cm]{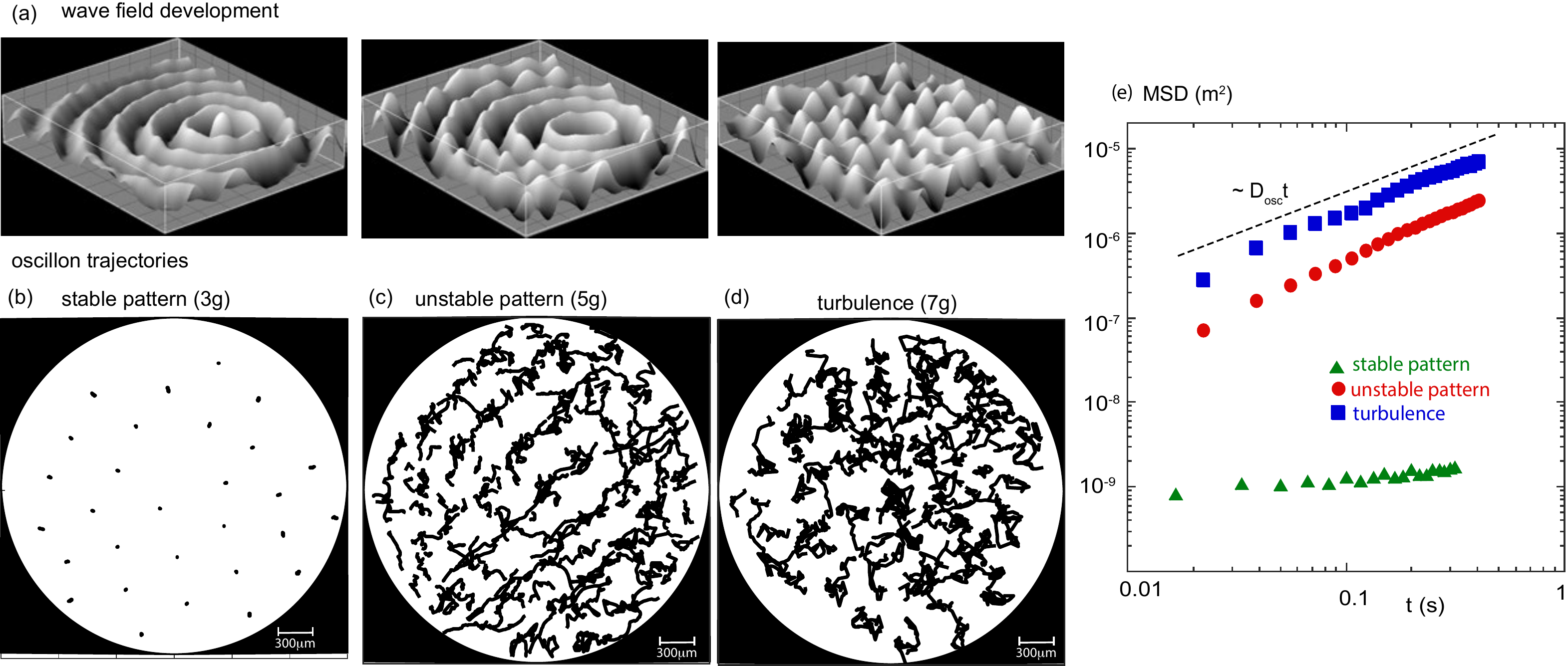}
\caption{\label{Fig_wave_mobility} (a) The development of the measured surface wave field in time, (b-d) trajectories of the wave maxima at different vertical accelerations 3g, 5g and 7g.  (e) MSD of the wave maxima at different vertical accelerations. }
\end{figure*}

The total mass of the biofilm within a microplate depends not only on the wave amplitude, but also on the stability of the wave pattern, Fig.~\ref{Fig_confocal}(a). The biomass is estimated from the measurements of the optical density (OD) of the crystal violet stain in the biofilm measured at the wavelength of 550 nm. The light absorption in the crystal violet solution (OD550) is compared with the control sample to obtain the normalised biofilm growth. Similarly, the OD at the wavelength of 600 nm is measured and compared with the control sample to evaluate the planktonic bacteria growth. For approximately constant density of the bacterial suspension, the OD of the crystal violet is noticeably (3-6 times) higher in the samples exposed to the surface waves in comparison with the control (non-vibrated) samples. The mass of the biofilm has a maximum at the vertical acceleration of $a=3g$. This corresponds to a reasonably intense waves still showing stable patterns. At higher acceleration, the mass is decreased in the presence of moving unstable pattern ($a=5g$), while for a turbulent wave field at $a=7g$, the biofilm mass is comparable to that in the control sample. In other words, the biofilm development is the strongest in the presence of stationary wave patterns, while the pattern mobility reduces the efficiency of the pattern formation and the biofilm

Measurements of the biofilm thickness are also performed using the Confocal Laser Scanning Microscopy (CLSM). Fig.~\ref{Fig_confocal}(b) shows the fluorescent image of the strongest biofilm corresponding to $a=3g$. CLSM images of two regions of interest indicate that at the thicker spot, the biofilm thickness is about 20-25 $\mu$m, Fig.~\ref{Fig_confocal}(c), while at the minimum it is a monolayer (Fig.~\ref{Fig_confocal}(d)) whose thickness is similar to that of the control (non-vibrated) sample: 2-5 $\mu$m. 

%The image of cross-sections of CLSM measurement in these two regions are shown in the Supplementary Fig. S1. Two videos of the CLSM at two different positions of the biofilm shows (1) attached bacteria in the middle of the biofilm (Supplementary Video 1), and (2) planktonic bacteria on top of the biofilms (Supplementary Video 2).

\begin{figure*}
\includegraphics[width=16cm]{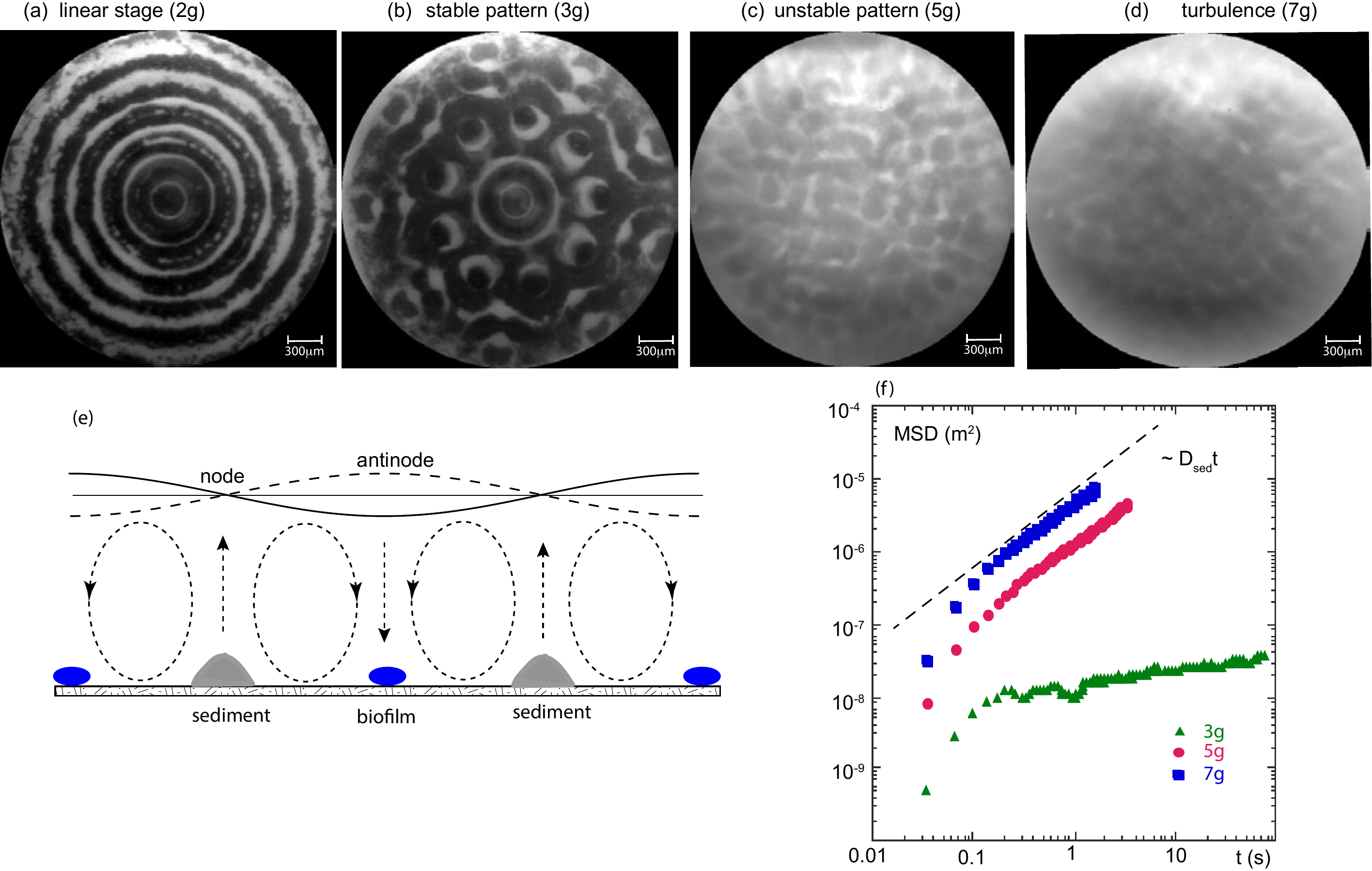}
\caption{\label{Fig_sediment_mobility} (a-d) TiO\textsubscript{2} sedimentation pattern at different vertical accelerations at 120Hz: 2g, 3g, 5g and 7g. (e) A schematic of the streaming pattern under standing surface waves \cite{Lighthill}. Blue and grey areas at the bottom indicate the locations of the biofilm growth and  TiO\textsubscript{2} sediment, respectively.  (f) The corresponding MSD of the sedimentation patterns at different vertical accelerations at 120 Hz.}
\end{figure*}

\subsection{Wave amplitude and horizontal mobility determine the biofilm growth}

Parametrically excited waves in vertically vibrated containers \cite{Douady1990} become disordered at higher vertical acceleration \cite{Goldman,Fineberg,ShatsPRL2012}. A wave in a circular cell develops at the first subharmonic of the vibration frequency $f_F=f_s/2$ above some acceleration threshold. At higher acceleration, the regular wave structure (concentric rings) is modulated by the cross-wave instability until the wave becomes broken into a wave field consisting of individual oscillating solitons, or oscillons \cite{Umbanhowar1996}. This process is illustrated in Fig.~\ref{Fig_wave_mobility}(a) during the first few seconds of the  development of the parametrically excited waves. The motion of individual oscillons becomes random, they chaotically move, collide and merge, as illustrated in the Supplementary Video 1. Such wave fields can be analysed by viewing oscillons as quasi-particles. It was shown that the diffusion coefficient characterising the random walk motion of the oscillons (derived from the mean-squared displacement of the maxima of the local surface elevation) is directly related to the fluid particle dispersion at the fluid surface \cite{Francois2015}. In a stable wave pattern, such as the one in Fig.~\ref{Fig_wave_biofilm}(b),  oscillons move very slowly around their equilibrium positions, but as the vertical acceleration is increased (Fig.~\ref{Fig_wave_biofilm}(c-d)), the horizontal motion of oscillons becomes essential for the wave dynamics. Figures.~\ref{Fig_wave_mobility}(b-d) show trajectories of oscillons in the horizontal plane for  three vertical accelerations, corresponding to the wave fields shown in Fig.~\ref{Fig_wave_biofim}(b-d). Figure~\ref{Fig_wave_mobility}(e) shows the mean-squared displacement (MSD) of oscillons as a function of time, $\langle \delta r^2 \rangle = \langle r(t)^2-r(0)^2\rangle $. At long times, the MSD is proportional to time which indicates the diffusive nature of the process: $\langle \delta r^2 \rangle = 2D_{osc}t$.  At the low vertical acceleration ($a=3g$), corresponding to the case of the stable pattern of Fig.~\ref{Fig_wave_biofilm}(b), the MSD is small, while at higher $a$ it is increased by up to four orders of magnitude, Fig.~\ref{Fig_wave_mobility}(e).

The diffusion coefficient $D_{osc}$ is a measure of the oscillon horizontal mobility and it is proportional to their root-mean-squared velocity $\langle \tilde{u}_{osc} \rangle_{rms}$. This velocity is linearly proportional to the rms velocity of the fluid particles at the fluid surface\cite{Francois2015}:  $\langle \tilde{u} \rangle_{rms} \approx 3\langle \tilde{u}_{osc} \rangle_{rms}$. In thin fluid layers, such as those in the reported experiments where the fluid thickness is less than the wave length ($h/\lambda \approx 0.4$), this fluid motion should also be sensed at the bottom of the fluid cell. We perform measurements of the motion of passive particles to compare with that of the active bacteria. Figure~\ref{Fig_sediment_mobility} shows patterns of the white powder of titanium dioxide (particles size of about 0.2 $\mu$m) sedimented at the bottom of the microplate. Patterns formed by the sedimented particles are similar to the patterns of biofilms, Fig.~\ref{Fig_wave_biofilm}(e-h). The main difference is that the passive TiO\textsubscript{2} particles are accumulated under the wave nodes, in contrast to the maxima of the biofilms which correspond to antinodes. 

The sedimentation of passive particles under waves is believed to be due to a wave streaming effect, or the generation of the time-averaged motion due to the rectification of the fluctuations of the fluid velocity (for example the Reynolds stress). The streaming motion was first considered by Rayleigh for acoustic waves \cite{Rayleigh,Lighthill} and it was later termed 'steady streaming' in incompressible flows \cite{Riley}.  Figure~\ref{Fig_2g}(c) illustrates the fast oscillating fluid motion induced by the wave. In addition to these oscillations, the surface wave induces a steady streaming motion as shown in Fig.~\ref{Fig_sediment_mobility}(e). The streaming moves fluid down from the antinodes and up at the nodal points. Such a streaming pattern has been confirmed experimentally in the Faraday waves \cite{Perinet2017}. In our experiments, the sedimentation of passive TiO\textsubscript{2} particles is also observed at the nodal points, in agreement with previous observations \cite{Saylor2005} and consistent with the streaming motion.

The increased horizontal mobility of the surface waves leaves a footprint on the sedimentation patterns. These patterns become blurry at higher accelerations $a=(5-7)g$. We track the sedimented particles near the bottom of the cell. The MSD of the sediment as a function of time is shown in Fig.~\ref{Fig_sediment_mobility}(f) for the vertical acceleration in the range of $a=(3-7)g$. The MSD is rather low in the presence of the stable wave pattern ($a=3g$), while in the presence of the unstable pattern ($a=5g$) and turbulence ($a=7g$), the diffusion coefficient $D_{sed}$ is higher by a few orders of magnitude. It was shown that on the surface of the fluid perturbed by steep Faraday waves, the horizontal diffusion coefficient is given by $D=\langle \tilde{u} \rangle_{rms}L_f$, where $L_f=\lambda_F/2$ is the forcing scale of the surface flow which is approximately equal to half the Faraday wave length. In our experiments $L_f \approx 2.5$ mm at the frequency of the parametrically excited waves of $f_F=60$ Hz. From the diffusion coefficient $D_{sed}$ of the sedimented particles of Fig.~\ref{Fig_sediment_mobility}(f), we estimate the rms particle velocity at the bottom $\langle \tilde{u}_{sed} \rangle_{rms}$ to be about $10^{-2}$ mm/s for the case of a stable pattern ($a=3g$) and about 1 mm/s in the turbulent regime at $a=7g$. The former velocity (10 $\mu$m/s) is less than the motility of active \textit{E. coli} bacteria which is in the range of 15 to 70 $\mu$m/s  \cite{Purcell1977,Guasto2012}. Thus the microorganisms can overcome the wave-induced fluid motion at lower wave amplitudes ($\le3g$), while in the turbulent regime, the wave-driven flow dominates over their motility.

\section{Summary and discussion}

Faraday waves in thin layers of media generate flows which affect the attachment of bacteria to the microplate bottom and the overall growth of biofilms. The patterns of mature biofilms (48 hours incubation time) reproduce patterns of the surface waves: the maximum thickness of biofilms is correlated with the locations of the peaks-troughs of the waves. The settlement of  passive TiO\textsubscript{2} particles, in contrast, occurs under the wave nodal points where the surface elevation does not change in time.

The sediment pattern of the TiO\textsubscript{2} particles agrees with the streaming patterns \cite{Rayleigh,Lighthill, Riley} and with recent experimental observations \cite{Perinet2017,Saylor2005}. To exclude the possibility that the size and the density of the sedimented TiO\textsubscript{2} particles play a role in their settlement locations, we performed experiments with inactive \textit{E. coli} bacteria in the suspensions of phosphate buffer saline with no nutrient. In such solutions, bacteria show substantially reduced motility \cite{Liao2003} and thus should sediment similarly to passive microparticles. Indeed, the settlement of inactive microorganisms is observed at the wave nodes (see Supplementary Fig. S1). This suggest that the ability of active \textit{E. coli} to overcome the wave-produced flows is an important factor which determines the selection of the attachment location at the bottom. When the fluid velocities exceed the swimming speed of the bacteria, no clear biofilm patterns are observed.

Other factors which determine the location of the biofilm, can be the wave-driven transport routes, delivering, for example, oxygen. It has been found that in the nutrient broth, \textit{E. coli} cells use oxygen very quickly and once oxygen is exhausted, the average velocity of the bacteria dramatically decreases \cite{Poon2016}. It is thus possible that the transport routes connecting the fluid surface with the bottom of the container, due to the wave streaming, is the main factor determining the locations of the biofilm development. 

To test how the streaming transport is affected by  the horizontal mobility of the waves, we visualise the mixing of the fluorescent dye (initially placed at the fluid surface) into the bulk (see Supplementary Information, Fig. S2.). In the presence of stable wave patterns ($a=(2-3)g$), the mixing occurs in the form of well defined stationary vertical plumes penetrating from the antinodes at the free surface to the bottom of the microplate. At higher accelerations, intense horizontal mixing of the fluid destroys vertical plumes and deterministic transport routes of, for example, oxygen from the fluid surface.  

\begin{figure}
\includegraphics[width=6cm]{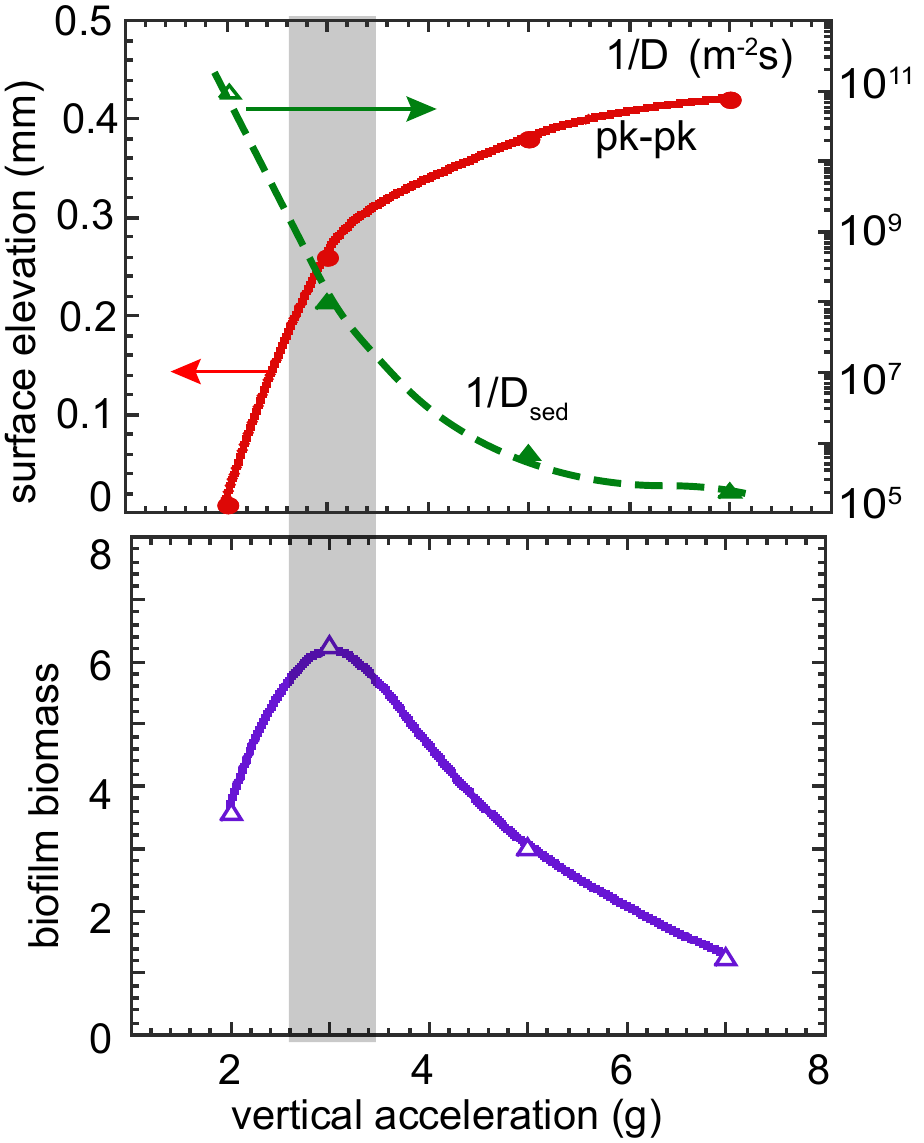}
\caption{\label{waveheight_diffusion} (a) Maximal surface elevation of the surface wave field (solid line ) and the inverse of the diffusion coefficient of the oscillon (dashed line), and (b) the total mass of attached biofilm, as a function of the vertical accelerations at 120Hz.}
\end{figure}

The maximum production of biofilm mass is observed at the intermediate wave intensity and at vertical accelerations corresponding to stable wave patterns and low horizontal transport of a fluid. Fig.~\ref{waveheight_diffusion}(a)  shows the peak-to-peak wave amplitude of the surface elevation and the inverse diffusion coefficient of the sedimented particles ($1/D_{sed}$) as a function of vertical acceleration. The strongest biofilm is observed at a finite wave height and relatively low horizontal mobility ($\propto D_{sed}$) of the microparticles near the bottom. In the fully turbulent flow at $a=7g$, no patterns are observed and the biofilm mass is the lowest.

The above results are scalable. By changing the frequency of the vertical vibration and the Faraday wave length in the range of $f_s=(45-120)$ Hz, we observe different spatial scales of the wave-induced patterns of biofilms (see Supplementary Information, Fig. S3.). The lower the $f_s$, the larger the scale of the biofilm patterns.

Presented results offer a simple yet efficient method of shaping biofilms on a solid substrate and allow increasing the biofilm development by inducing  waves on the surface of the media.

\section{Methods}

\textit{Experimental setup}. The sample holders  housed in a temperature-controlled incubator (37 $^{\circ} C$), are vertically vibrated by an electrodynamic shaker. The vertical acceleration of the microplates is accurately monitored. The frequency of the vertical vibration can be changed in the range of (0-1.2) kHz, the maximum acceleration is up to $20g$. Most of the results presented in this paper are conducted at 120 Hz, with vertical acceleration from $2g$ to $7g$, which corresponds to a vertical displacement of 0.07 mm ($2g$) to 0.24 mm ($7g$). Experiments are also conducted at $f_s=60$ Hz at $a=0.8g$ (0.11 mm vertical displacement) to $3g$ (0.4 mm displacement). Polystyrene microparticles (1 $\mu$m) and TiO\textsubscript{2} particles (200 nm) are used as tracers in the sedimentation studies. The images of the sediment are captured using Andor Zyla camera mounted above the fluid plates. 

A synthetic Schlieren technique developed in Ref. \cite{Moisy2009} is used to measure wave fields on the water surface. The method is based on the analysis of the refracted image (above the surface) of a random dot pattern placed under the transparent bottom of the fluid plate. When the surface is flat, a reference image is obtained. The apparent displacement field between the refracted image and the reference image is determined, which is then used for the reconstruction of the instantaneous surface elevation measurements. The surface elevation is used to identify and track the horizontal motion of the oscillons. After preprocessing the images using ImageJ software, the isolines of the surface elevation are analyzed to identify the oscillons. A particle tracking algorithm is applied to follow trajectories of the oscillons using a nearest neighbor algorithm.

\textit{Bacteria culture}. Microplates  (six-well) containing 2 ml of media (Tryptic Soy Broth $ + 100 { \mu g} /{ml}$ ampicillin) are inoculated with 20 $\mu l$ of overnight culture of \textit{Escherichia coli} GFP (ATCC 25922 GFP) diluted to 0.1 OD. The microplates are vibrated for 24 hours. The samples are further incubated for another 24 hours at 37 $^{\circ} C$. The control samples are incubated for 48 hours in a non-vibrated incubator. At least six measurements are performed each time for two overnight growths, each repeated three times. At the end of each experiment, the top solution from the microplates is collected to measure the OD600 to characterise the planktonic bacterial growth. The attached bacteria and biofilms are washed before they are stained using 0.1 $\%$ crystal violet (CV) and incubated for 20 minutes. After the development of the stain, the samples are washed several times to remove unabsorbed CV. Then the absorbed CV is dissolved in 2 ml of ethanol solution (20 $\%$ acetone and 80 $\%$ ethanol) to measure the OD of CV absorbed by the attached biomass. All optical density measurements are performed using a Varioskan Lux multimode reader. 3D structures and movies of biofilms were obtained using an upright Zeiss LSM780 UV-NLO confocal microscope.

\acknowledgments{
This work was supported by the Australian Research Council Discovery Projects and Linkage Projects funding schemes (DP160100863, DP190100406 and LP160100477). H.X. acknowledges support from the Australian Research Council's Future Fellowship (FT140100067). N.F. acknowledges support by the Australian Research Council's DECRA award (DE160100742).The authors acknowledge the technical assistance of Centre for Advanced Microscopy (ANU).

Author contributions:

H.X and M.S designed the project. S.H.H and H.X. conducted the experiments. J.B.G. conducted the wave measurements. H.P. designed the experimental setup. N. F. analyzed the sedimentation of passive particles. H.X and M.S wrote the paper. All authors reviewed the paper. 
}

%\bibliographystyle{aipnum4-1}
%\bibliography{biofilm_draft_2019}% Produces the bibliography via BibTeX.

\end{document}